\documentstyle[aps,prl,multicol,graphicx]{revtex}
\include{psfig}

\begin{document}
\author{Francesco Sciortino, Piero Tartaglia and Emanuela Zaccarelli}
\address{Dipartimento di Fisica, INFM Udr and Center for Statistical Mechanics and Complexity, \\ 
Universit\`{a} di Roma La Sapienza, I-00185 Rome. Italy.}

\title{Evidence of a higher-order singularity in dense short-ranged attractive colloids}
\date{\today}
\maketitle

\begin{abstract}
We study a model in which particles interact
through a hard-core repulsion complemented by a short-ranged
attractive potential, of the kind found in colloidal suspensions.
Combining theoretical and numerical work we locate the
line of higher-order glass transition singularities and its end-point
--- named $A_4$ --- on the fluid-glass line. Close to the $A_4$ point,  
we detect logarithmic decay of density correlations and sub linear power-law increase of the mean square displacement, for time intervals up to four order of magnitudes. We establish the presence of the
$A_4$ singularity
by studying how the range of the potential affects the time-window where anomalous dynamics is observed.
\end{abstract}



\begin{multicols}{2}
\smallskip%
Recently, a great deal of interest has grown around dynamical
phenomena arising in dense colloidal suspensions when particles
interact through a hard-core repulsion followed by a rather
short-ranged attractive
potential\cite{sciortino02,frenkelnature,articlesinroyal}.  Some of
these unusual phenomena, initially predicted
theoretically\cite{fabbian99,bergenholtz99,dawson00}, have been
observed experimentally\cite{mallamace00,malla02,pham02,eckert02} and
in numerical
simulations\cite{puertas02,puertas03,foffi02,zaccarelli02,newzacca}.
The novelty, as compared to the well-studied hard-sphere-colloids
case, comes from the possibility of generating structurally arrested
states with an additional physical mechanism, driven by the range of
the attractive interaction. Indeed, in the case of hard-sphere,
structural arrest is controlled by the well-known excluded volume cage
effect, where particle motion is hindered by the presence of
neighboring ones. This mechanism generates a localization length of
the order of 10 \% of the particle diameter $\sigma$. In the
additional mechanism, the motion of the particles is restricted by the
adhesiveness, i.e. the formation of non-permanent bonds due to the
attractive part of the inter-particle potential; particles are
confined by bonds and the corresponding localization length is fixed
by the attraction range $\Delta$.  When $\Delta \ll \sigma$, the
interplay of the two localization mechanisms creates liquid states for
packing fractions higher than those possible for a pure hard-sphere
system. This means that the attraction stabilizes the liquid and leads
to a reentry phenomenon, where
this new liquid state can
arrest into a glass by cooling as well as heating. The experimental
and numerical confirmation of this
prediction\cite{pham02,eckert02,puertas02,puertas03,foffi02,zaccarelli02}
suggests that the mode-coupling theory (MCT) \cite{goetze91}, which
was the basis of the cited theoretical work
\cite{fabbian99,bergenholtz99,dawson00}, can contribute to explain the
slow dynamics in short-ranged attractive colloids.

\begin{figure}
\begin{center} 
\includegraphics[width=8.5cm,angle=0.,clip]{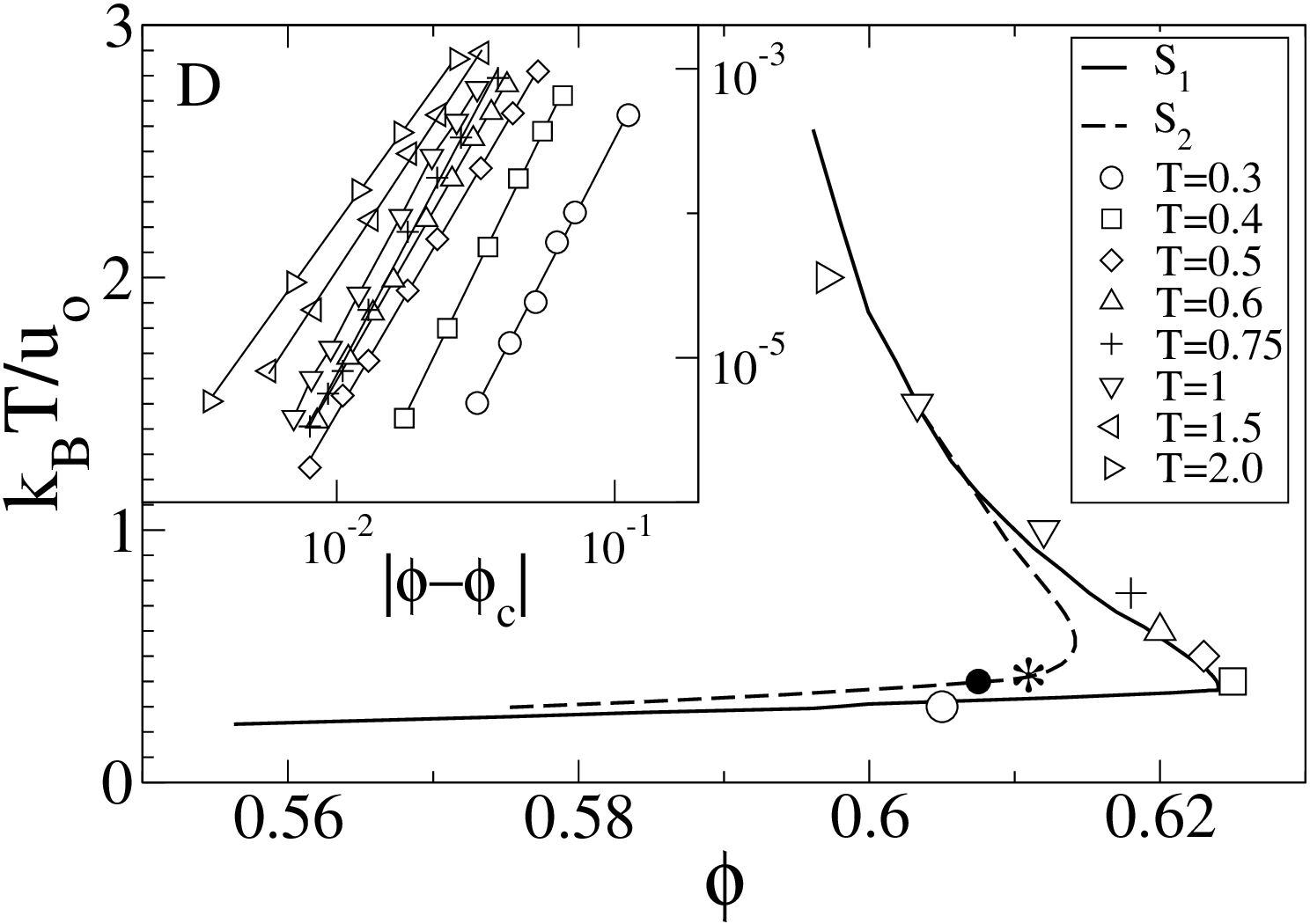}
\caption{Glass transition line for the system $S_1$ calculated  through the vanishing of the diffusion (open symbols) and calculated with MCT(full line). The MCT ideal glass line for  $S_2$ is given by the dashed line. The star indicates the location of the $A_4$ singularity (on the $S_2$ ideal glass line) while a filled square locates the $A_3$ point (on the $S_1$ line).
The MCT curves have been transformed according to the mapping discussed in the text. The $T$ and $\phi$ of the simulated point are indicated by the black dot. The inset shows the power-law fits for the isothermal diffusion coefficient $D$ (data from Ref.\protect\cite{zaccarelli02}) for  $S_1$  as a function of  $\phi - \phi_c$ (see text).
}
\label{fig:phase}
\end{center}
\end{figure}

These systems are characterized by three control parameters --- the packing fraction $\phi$, the ratio $\theta$ of the thermal energy $k_BT$ to the typical well depth $u_o$  and the range $\Delta$ of the attractive potential. Within MCT, the phase diagram of this three-dimensional control parameter space (shown in Fig.\ref{fig:phase}) is organized around a critical point $(\phi^*, \theta^*, \Delta^*)$, referred to as a type $A_4$ higher-order glass transition singularity in MCT classification. 
$A_4$ is the end-point of a line of higher-order singularities (of type $A_3$). From a physical point of view,   $A_4$ is characterized  for being the only singularity point accessible from the liquid phase. For $\Delta > \Delta^*$ no singular points are predicted by the theory, while for $\Delta < \Delta^*$
the $A_3$ singularity points are buried in the glass phase and their presence can be only observed indirectly\cite{mallamace00,pham02,puertas02,zaccarelli02}.
Near the $A_4$ singularity, MCT predicts a structural relaxation dynamics, which is utterly different from that known for conventional glass-forming liquids. It is ruled by logarithmic variations of correlators as function of time and subdiffusive increase of the  mean squared displacement (MSD).  Theory makes precise predictions for the  time-interval where  this unusual dynamics  is expected, as well as for its variation  with changes of the 
$\phi$, $\theta$ and $\Delta$ parameters.  Indications of logarithmic dynamics have been reported\cite{mallamace00,pham02,puertas02,zaccarelli02},
but neither a systematic study of the dependence on the control parameters
has been published nor the $A_4$ point has been located numerically. It is in fact difficult to discriminate between straight lines caused by some inflection points in the usual logarithmic representation of correlators  and the logarithmic decay laws generated by the higher-order singularity. 
In this Letter, we establish the existence of the location of the $A_4$ point for a well defined model, amenable of simultaneous numerical and theoretical treatment. Reported data --- with control parameters explicitly chosen close to the $A_4$ point --- exhibit the mentioned laws over time intervals up to four orders of magnitude. More importantly, we show that the decay patterns vary with changes of the control parameters and wave-vectors as expected\cite{sperl02,sperl03}, properly testing the theoretical predictions.

We simulate a $50\%-50\%$ binary mixture of $N=700$ hard spheres of
mass $m$, with diameters $\sigma_{_{AA}}$ and $\sigma_{_{BB}}$ and
ratio $\sigma_{_{AA}}/\sigma_{_{BB}}=1.2$. The hard core between
particles of different type $\sigma_{_{AB}} = 0.5
(\sigma_{_{AA}}+\sigma_{_{BB}})$. The hard core potential is
complemented by an attractive square well potential of depth $u_o$,
independent on the particle type\cite{units}.  The small asymmetry in
the diameters is sufficient to prevent crystallization at high values
of $\phi$.  We focus on two specific systems, which we label $S_1$ and
$S_2$, differing in the width of the square well $\Delta_{ij}$.  $S_1$
and $S_2$ have $\Delta_{ij}=0.031 \sigma_{ij}$, and
$\Delta_{ij}=0.043\sigma_{ij}$ respectively ($i=A,B$).  The system
$S_1$ has been extensively studied in previous
simulations\cite{zaccarelli02}. As discussed in the following, $S_2$
is chosen to coincide with the critical amplitude parameter within the
accuracy of our calculations. Averages over five independent
realizations have been performed to reduce noise. Equilibration has
been carefully checked.  We note that our results refer to Newtonian
dynamics, but are relevant also to Brownian dynamics in the structural
relaxation time window \cite{franosch}.  

To estimate the predicted location of the $A_4$ point for the binary mixture square well model  (and consequently select the parameters 
to be used in the simulations) 
we proceed as follows. First, we calculate the glass line $\phi_c(T)$  for the  mixture $S_1$  extrapolating diffusivity $D$ data\cite{zaccarelli02} according to $D \sim (\phi -\phi_c(T))^{\gamma(T)}$, for eight different $T$.  Second, we calculate the MCT
ideal glass transition line. The partial structure factors,  the only input needed,  are calculated with the Percus-Yevick approximation, numerically solving the Ornstein-Zernike equation\cite{mct-calculations}.  Third, we map the theoretical glass line on the simulation line, following the procedure first used by Sperl\cite{sperl03}.  Indeed  MCT cannot reproduce the numerical values for $\phi_c$.  As shown in Fig.\ref{fig:phase}, the linear 
transformation $\phi \rightarrow 1.897\ \phi -0.3922$ and $T \rightarrow 0.5882\ T - 0.225$, allows to superimpose the MCT result with the simulation data, in the studied region of $T$ and $\phi$.  
By calculating the MCT ideal glass transition line for several values of $\Delta$, the theoretical location of the $A_4$ point for the binary mixture square-well model  is obtained.  Assuming that, to a first approximation, the same linear transformation holds also for the critical value of the well width, we find that  the $A_4$ location
maps to  $\Delta^*_{ij}=0.043 \sigma_{ij}$,  $\phi^*=0.611$ and $\theta^*=0.416$.  This allows us to
select a state point with parameters close to the critical ones 
and perform a simulation (in thermodynamic equilibrium) close to $A_4$.
We choose\cite{comment} $\theta=0.4$ and $\phi=0.6075$ and  compare,   for the same $T$ and $\phi$,  the dynamics for $\Delta^*_{ij}$ (system $S_2$) and for the close-by value $\Delta_{ij}=0.031 \sigma_{ij}$ ($S_1$).

\begin{figure}
\begin{center} 
\includegraphics[width=9cm,angle=0.,clip]{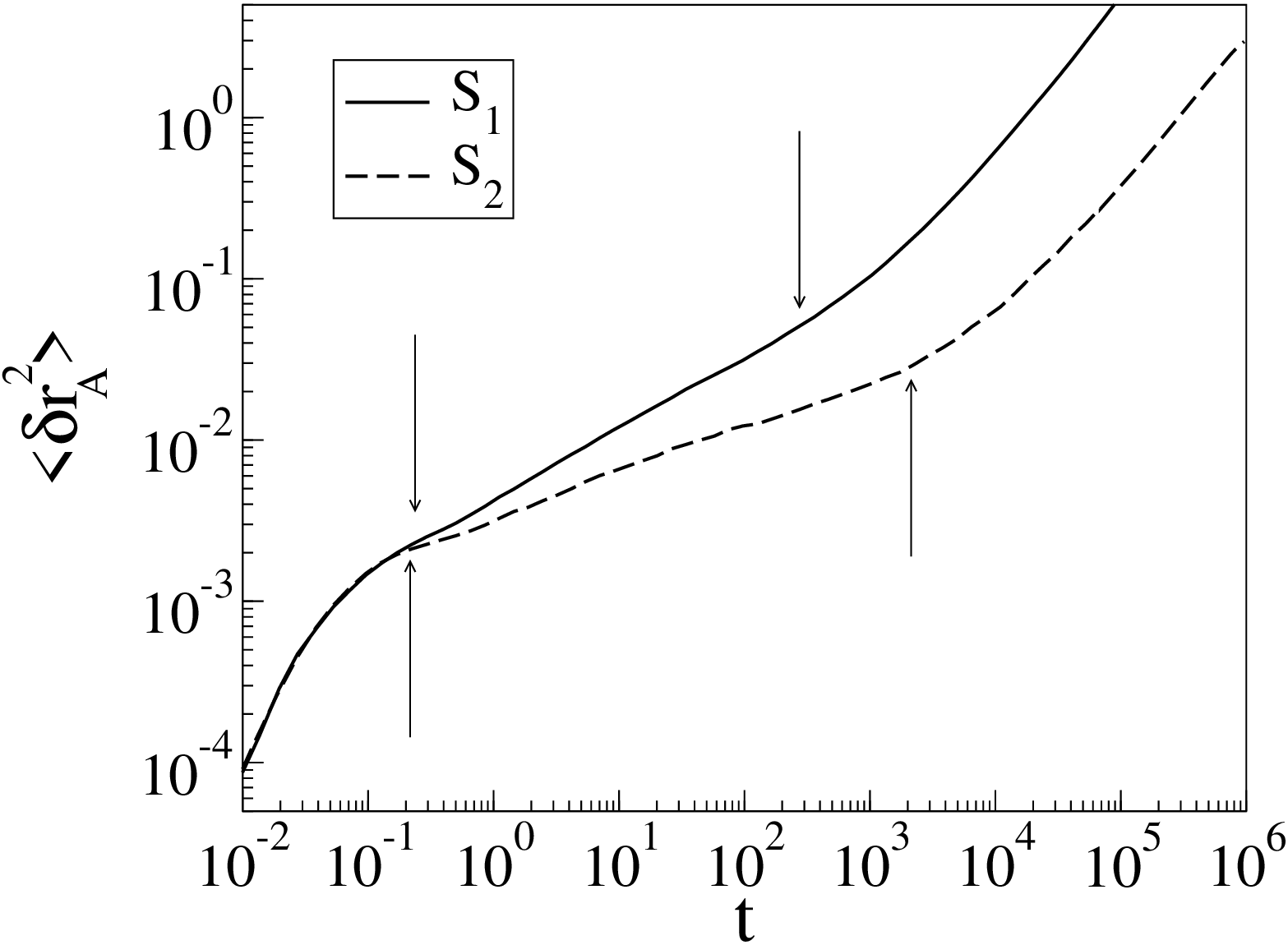}
\caption{ Mean squared displacement of the $A$ particles,
 $\langle \delta r_A^2\rangle $,  showing sub-diffusive behaviour within the marked intervals. 
At long times, the diffusive behavior $\langle \delta r_A^2\rangle \sim t$  is recovered.}
\label{fig:msd}
\end{center}
\end{figure}

\begin{figure}
\begin{center} 
\includegraphics[width=8cm,angle=0.,clip]{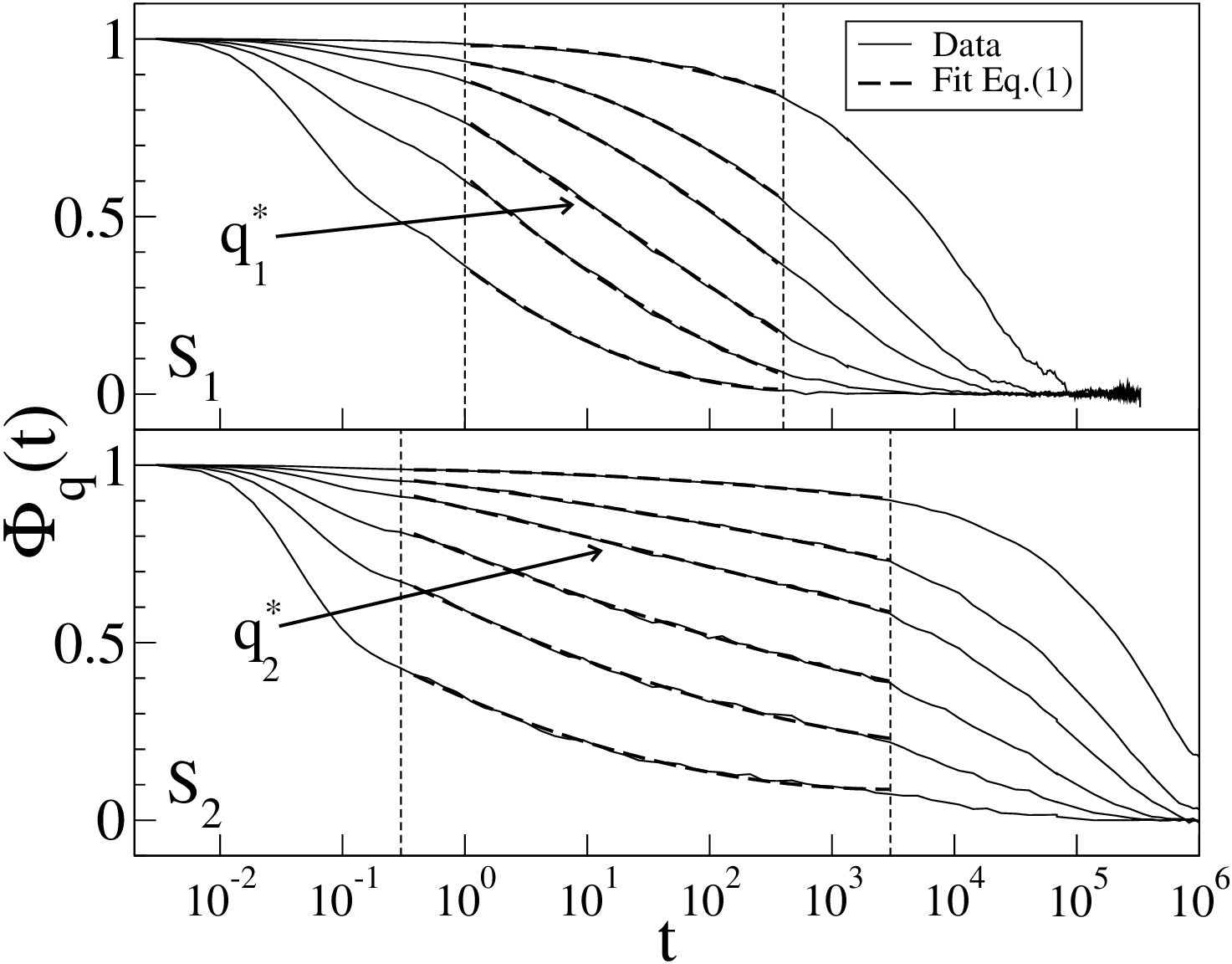}
\caption{$A$-type particle density correlators for  
$S_1$ (top panel) and $S_2$ (bottom panel), 
showing six different wave vectors (from top to bottom  $q\sigma_{_{BB}}=6.7, 11.7, 16.8, 23.5, 33.5$ and $50.3$).  For  correlators 
indicated by an arrow, a logarithmic behavior
is observed within the selected time window.  The dashed lines show the fits according to  Eq.\ref{eq:log}. The vertical dashed lines indicate the fitting interval.}
\label{fig:correlators2}
\end{center}
\end{figure}

Fig.\ref{fig:msd} shows the MSD for the larger species $\langle \delta r_A^2\rangle$ for both systems. We notice the presence of a subdiffusive regime at intermediate times, i.e. a variation according to $ln(\langle \delta r_A^2\rangle)=f+a\ ln(t/\tau)$ for about three decades for $S_1$, where $a_1=0.44$. The logarithmic regime extends by more than a decade if the attraction range is increased to that for system $S_2$. Simultaneously the exponent $a$ decreases to $a_2=0.28$. 
Fig.\ \ref{fig:correlators2} shows the density-fluctuation auto-correlation functions $\Phi_q(t)$  for a representative set of wave vectors $q$.  
The decay curves  in  Fig.\ \ref{fig:correlators2} do not show the two steps scenario with a plateau characteristic of conventional glass forming liquids.   Indeed it  is impossible to fit these curves for large $q$ with the standard stretched exponential function. Instead, there is a  region of clear logarithmic decay at $q_1^*\sigma_{_{BB}}=23.5$  for $S_1$ and
at  $q_2^*\sigma_{_{BB}}=16.8$ for $S_2$. The time-intervals over which logarithmic decay is observed  are of similar size as those for the MSD. For $q<q^*$, the $\Phi_q$ vs $ln\ t$ curves are concave, and for $q > q^*$  convex.

To show that the above described features are consistent with the ones predicted by MCT, we cite the general asymptotic decay law for the correlation function of a generic variable $X$,  near the higher-order singularity\cite{sperl02}
\begin{equation}
\label{eq:log}
\Phi_X(t)=f_X-h_X \left[B^{(1)} ln(t/\tau) + B^{(2)}_X ln^2(t/\tau)\right].
\end{equation}
Here $\tau$ abbreviates a time-scale which diverges if the state approaches the singularity. The formula is obtained by asymptotic solution of the MCT equations, using the parameter differences $\phi^*-\phi$, $\theta^*-\theta$, $\Delta^*-\Delta$ as small quantities, say, of order $\epsilon$. The
coefficient $B^{(1)}$ is of order $\sqrt\epsilon$ while
$B^{(2)}$ is of order $\epsilon$. The amplitude $h_X$ is independent of $\epsilon$. The first term $f_X$  is the sum of the non-ergodicity parameter of variable $X$ at the singularity and a correction of order $\epsilon$. Terms of order $\epsilon^{3/2}$  are neglected\cite{sperl02}.
We can interpret  $ln\langle \delta r^2_A\rangle$ also 
 according to Eq. (\ref{eq:log}). Hence, the straight line for $S_1$ with slope $a_1$ shown in Fig. \ref{fig:msd} is consistent with the assumption that $S_1$ is close to the singularity.
Changing from $S_1$ to $S_2$, the slope $a=h_{_{MSD}}B^{(1)}$ has to decrease and the range of validity of the leading order description has to expand. This is demonstrated impressively by the data.

\begin{figure}
\begin{center} 
\includegraphics[width=9cm,angle=0.,clip]{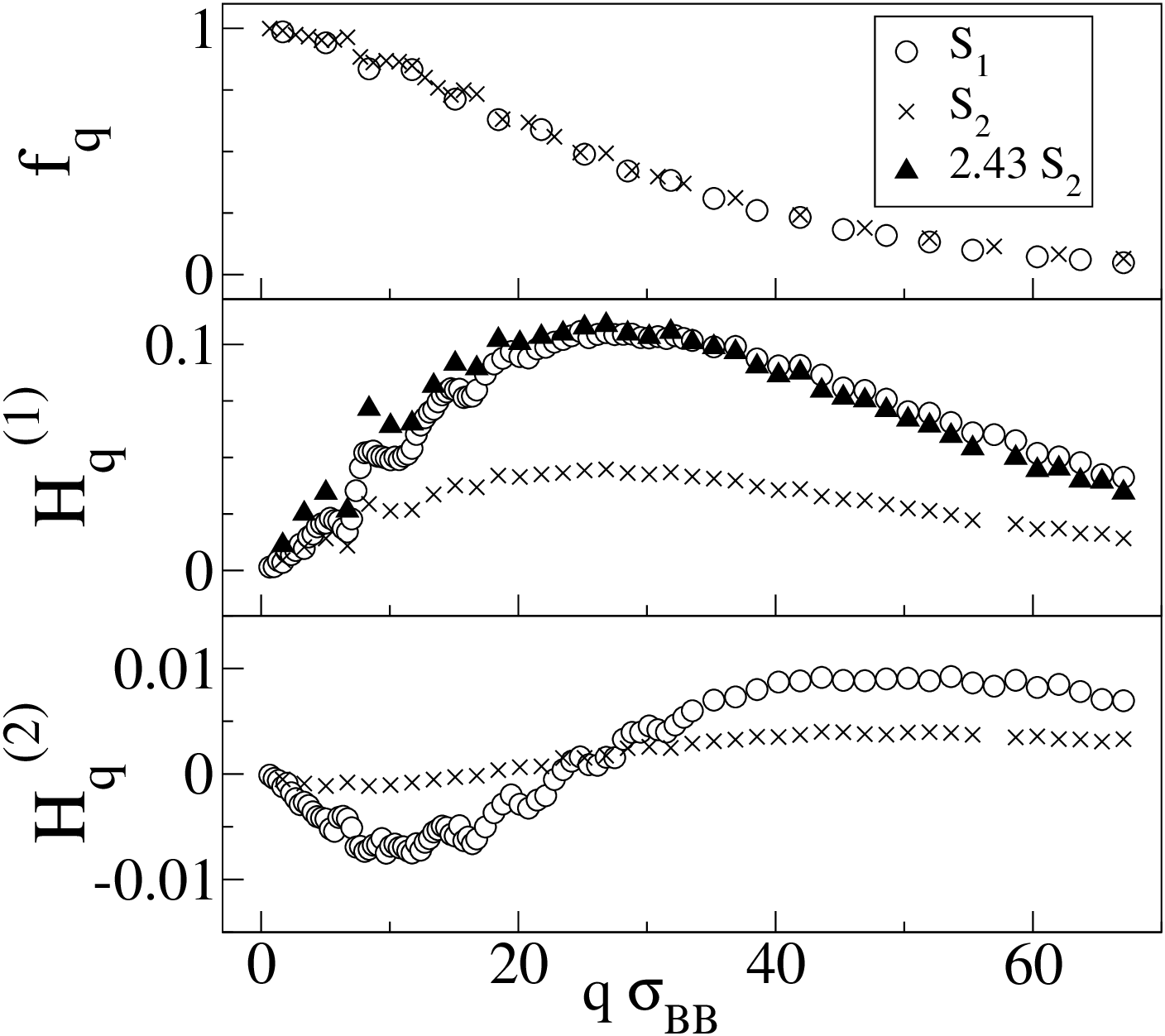}
\caption{The fitting parameters $f_q$,  $H_q^{(1)}$ and $H_q^{(2)}$ of the asymptotic logarithmic law, for $S_1$ and $S_2$.   The  central panel also shows that
a multiplication by 2.43 of    $H_q^{(1)}$ for the $S_2$  system
gives  $H_q^{(1)}$ for $S_1$, confirming the
factorization of $H_q^{(1)}$. 
At the wave vector where $H_q^{(2)}=0$  ($q^*_2\sigma_{_{BB}}=23.5$ for $S_1$ and $q^*_2\sigma_{_{BB}}=16.8$ for $S_2$) 
correlation functions display a pure logarithmic decay.
}
\label{fig:hq}
\end{center}
\end{figure}

To estimate the possibility of describing study the time dependence of
$\Phi_q(t)$ according to Eq.\ \ref{eq:log}, we fit the density
autocorrelation functions to a quadratic polynomial in $log(t/\tau)$
for different $q$ values. Fits
are reported in Fig.\ \ref{fig:correlators2}. 
The fitting time window extends from about $2.5$ decades for the state
point $S_1$ to four decades for $S_2$.  The fitting parameter $f_q$,
shown in Fig.\ref{fig:hq}, provides an estimate of the non-ergodicity
parameter at the $A_4$ point.  We find that $f_q$ does not depend on
the state point. This confirms the preceding conclusion that the
studied state points are very close to the singularity and that the
order $\epsilon$ correction in $f_q$ cannot be detected.

The fit parameters for the coefficient $H^{(1)}_q \equiv h_q B^{(1)}$
are reported in the middle panel of Fig.~\ref{fig:hq}. The decrease of
$H^{(1)}_q$ upon changing from $S_1$ to $S_2$ is in agreement with the
prediction that $B^{(1)}$ tends to zero with $\epsilon$ approaching
zero.  As shown in Eq.\ref{eq:log}, $H^{(1)}_q$ factorizes into a
control-parameter independent factor $h_q$, depending on $q$, and a
control-parameter dependent factor $B^{(1)}$, independent of $q$.
This implies that the $q$-dependence of $H^{(1)}_q$ should be the same
for $S_1$ and $S_2$.  As shown in Fig.\ref{fig:hq}, this property is
verified by the data. The same property does not apply to $H^{(2)}_q
\equiv h_q B^{(2)}_q$ because of the $q$-dependence of $B^{(2)}_q$.
Moreover, the fitting results confirm that $H^{(2)}_q$ is smaller than
$H^{(1)}_q $, as expected being the first of order $\epsilon$ and the
second of order $\sqrt\epsilon$.  The wave-vector value where
$B^{(2)}_q=0 $ allows to identify the characteristic length scale
associated to the pure logarithmic decay (see
Fig.\ref{fig:correlators2}). Such length scales are much shorter than
the typical first neighbor shell.

\begin{figure}
\begin{center} 
\includegraphics[width=8cm,angle=0.,clip]{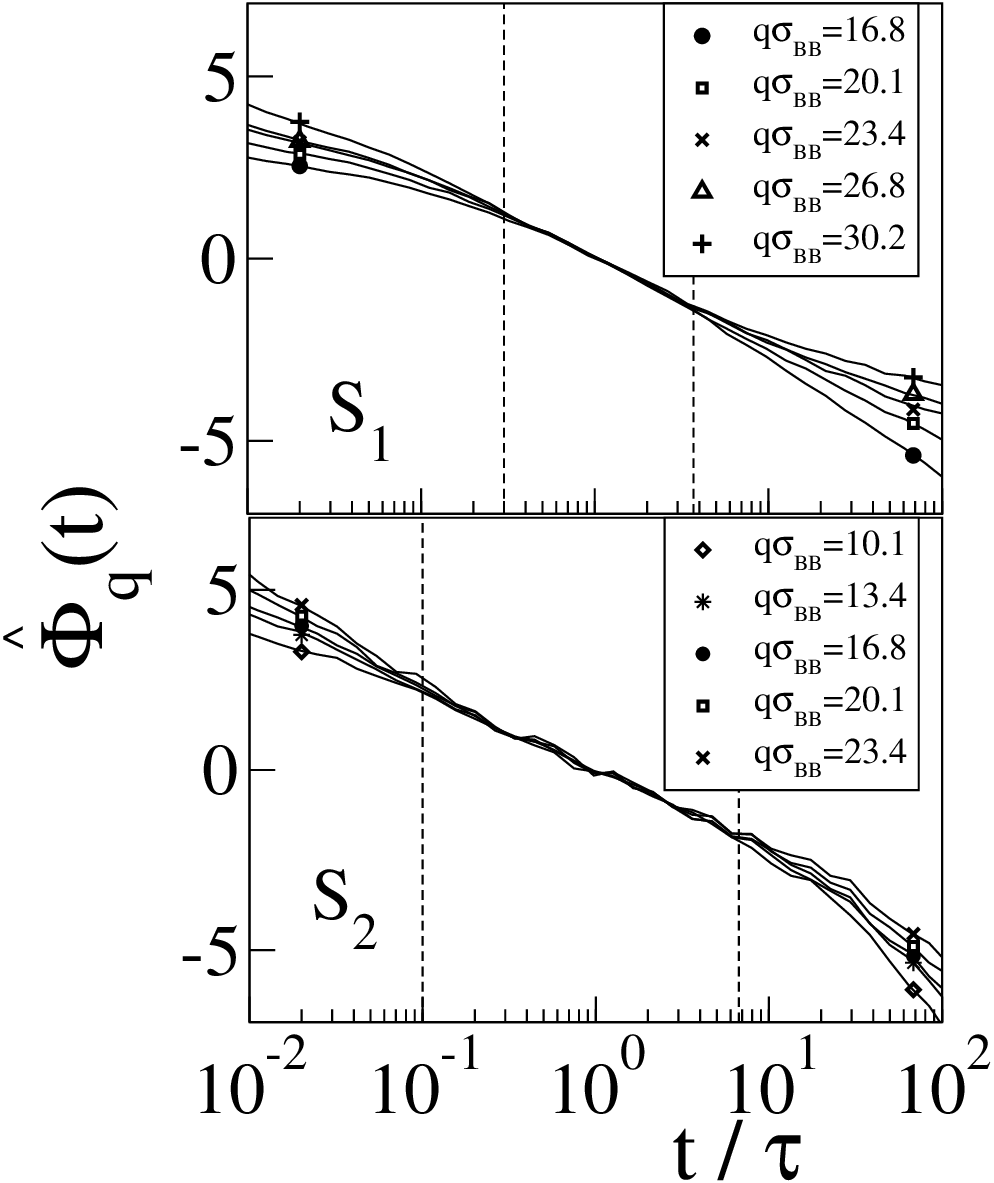}
\caption{ Scaled correlation functions $\hat{\Phi}_q(t)$ (see text) vs $ln(t/\tau)$ for the two samples and selected values of $q\sigma_B$ around the characteristic values  $q^*$ for which $H^{(2)}_q$ is close to zero. The vertical dashed lines indicate the time interval over which the scaling is observed. The values of $\tau$ are $10$ and $10^{2}$ for $S_1$ and $S_2$ respectively.
}
\label{fig:correlators3}
\end{center}
\end{figure}

To explicitly test the mentioned factorization,
Fig.~\ref{fig:correlators3} shows a set of correlators rescaled to
$\hat{\Phi}_q(t)=(\Phi_q(t)-f_q)/H^{(1)}_q$. MCT predicts that in
leading order $\epsilon$ all $\hat{\Phi}_q(t)$ vs $ln(t)$ curves
collapse on the logarithmic decay law $-ln(t/\tau)$. This is indeed
shown  for a large time interval that expands upon approaching the
singularity, as it is demonstrated by comparing  results for
$S_1$ and $S_2$\cite{notamct}.

Results reported in this Letter have an intrinsic value associated to
the observation, in a particularly simple system --- a square well
potential ---, of a particularly complex dynamics over more than four
decades in time.  They show that MCT, a theory essentially developed
to address the problem of the excluded volume glass transition is able
--- without any modification --- to handle the logarithmic dynamics
and to provide an interpretative scheme in term of $A_4$ point .  The
numerical results and the comparison with the theoretical predictions
do constitute, in fact, a stringent verification that the logarithmic
dynamics in the density autocorrelation functions and the subdiffusive
behavior in the MSD can be fully rationalized by MCT. Indeed, data in
Figs.\ref{fig:correlators2} and \ref{fig:hq} are in agreement with MCT
predictions (Fig. 12 and 13 of Ref.\cite{sperl03}). The simplicity of
the model and the complexity of the dynamics suggest that the
short-range attractive colloids have the potentiality to become a
benchmark for the development of extended theories of the glass
transition.


We acknowledge support from MIUR PRIN and FIRB and
INFM PRA-GENFDT. We thank W. G\"otze for illuminating discussions, G.~Foffi and M. Sperl for comments  and S.~Buldyrev for providing us the molecular dynamics code for square-well binary systems.

\end{multicols}


\begin{references}

\bibitem{sciortino02}
F. Sciortino,
Nature Materials, News and Views, {\bf 1}, 145 (2002).

\bibitem{frenkelnature} 
D. Frenkel, Science {\bf 296}, 106 (2002)


\bibitem{articlesinroyal} Articles in Faraday Discuss. {\bf 123} (2003).
 

\bibitem{fabbian99}  
L. Fabbian,  
W. G\"{o}tze, F. Sciortino, P. Tartaglia and F. Thiery, 
Phys. Rev. E {\bf 59}, R1347 (1999) and Phys. Rev. E {\bf 60}, 2430 (1999).

\bibitem{bergenholtz99}  
J. Bergenholtz and M. Fuchs, 
Phys. Rev. E {\bf 59}, 5706 (1999).

\bibitem{dawson00}  
K. A. Dawson,  G. Foffi, M. Fuchs, W. G\"otze, F. Sciortino, M. Sperl,
P. Tartaglia, Th. Voigtmann, E. Zaccarelli 
Phys. Rev. E {\bf 63}, 11401 (2000).

\bibitem{mallamace00}  
F. Mallamace,
P. Gambadauro, N. Micali, P. Tartaglia, C. Liao and S. H. Chen, 
Phys. Rev. Lett. {\bf 84}, 5431 (2000).

\bibitem{malla02}
W.R. Chen, S.H. Chen and F. Mallamace, Phys. Rev. E, {\bf 66} 021403 (2002).

\bibitem{pham02} 
K.N. Pham, A.M. Puertas, J. Bergenholtz, S.U. Egelhaaf,
A. Moussaid, P.N. Pusey, A.B. Schofield, M.E. Cates, M. Fuchs 
        and W.C. Poon 
Science {\bf 296}, 104 (2002).

\bibitem{eckert02} 
T. Eckert and E. Bartsch, 
Phys. Rev. Lett. {\bf 89}, 125701 (2002).

\bibitem{puertas02}
A. M. Puertas, M. Fuchs, M. E. Cates, 
Phys. Rev. Lett. {\bf 88}, 098301 (2002) 

\bibitem{puertas03}
A. M. Puertas, M. Fuchs, M. E. Cates,
Phys. Rev. E {\bf 67}, 031406 (2003).

\bibitem{foffi02} 
G. Foffi, 
K.A. Dawson, S. Buldyrev, F. Sciortino, E. Zaccarelli and P. Tartaglia,
Phys. Rev. E  {\bf 65}, 050802  (2002).

\bibitem{zaccarelli02}
E. Zaccarelli  
G. Foffi, K. A. Dawson, S. V. Buldyrev, F. Sciortino, P. Tartaglia,
Phys. Rev. E  {\bf 66}, 041402 (2002).

\bibitem{goetze91}  
W. G\"{o}tze, 
in {\it Liquids, Freezing and the Glass
Transition}, edited by J. P. Hansen, D. Levesque, and J. Zinn-Justin 
(North Holland, Amsterdam, 1991).


\bibitem{sperl02}  
W. G\"{o}tze and M. Sperl, 
Phys. Rev. E {\bf 66}, 011405 (2002).

\bibitem{units} Distances are in units of $\sigma_B$, energy and $T$
in units of $u_0$ ($k_B=1$). Time is in units of
$\sigma_B\cdot(m/u_0)^{1/2}$, mass of $m$.

\bibitem{mct-calculations}  We have solved  the 
Ornstein-Zernike equation on a grid of 20000 $q$ values,
with mesh 0.3141593 and the MCT equations on a grid of 2000  $q$ with the same mesh.  

\bibitem{sperl03}  
M. Sperl,  Phys. Rev. E {\bf 68} 031405 (2003).

\bibitem{comment} Equilibrium simulations at state points closer to
the $A_4$ point are not feasible with the present computational
resources. Present data requested several months of cpu time on a
beowulf Athlon cluster.

\bibitem{franosch} T. Franosch, W. G\"otze, M. R. Mayr and
A. P. Singh, J. Phys. Condens. Matter {\bf 71}, 235 (1998).

\bibitem{notamct} All curves should have the same tangent for $t=\tau$
but, to order $\epsilon$, be either convex or concave depending on the
sign of $B^{(2)}_q $. This important signature of a higher-order
glass-transition singularity is exhibited by the data.

\bibitem{newzacca} E. Zaccarelli, G. Foffi, F. Sciortino and P. Tartaglia,
Phys. Rev. Letts. {\bf  91}, 108301 (2003).



\end{references}
\end{document}